\documentclass[preprint]{elsarticle}
\usepackage{amsmath, amsthm, amssymb}
\usepackage{subfig}
\usepackage{rotating}
\usepackage{verbatim}
\usepackage{dcolumn}
\newcolumntype{d}{D{.}{.}{2.5}}
\newcolumntype{s}{D{.}{.}{1.2}}

\newcommand{\comments}[1]{}

\def\bP{{\bf P}}
\def\bR{{\bf R}}
\def\avg#1{\langle #1\rangle }

\journal{Solid State Sciences}

\begin{document}

\begin{frontmatter}

\title{Phase Transitions of Boron Carbide: Pair Interaction Model of High Carbon Limit}

\author[cmu]{Sanxi Yao}
\ead{sanxiy@andrew.cmu.edu}
\author[cmu,duke]{W. P. Huhn}
\ead{wph@andrew.cmu.edu}
\author[cmu]{M. Widom\corref{cor1}}
\ead{widom@andrew.cmu.edu}

\cortext[cor1]{Corresponding Author}
\address[cmu]{Department of Physics, Carnegie Mellon University, 5000 Forbes Ave, Pittsburgh, PA, 15232, United States of America, 412-268-7645}
\address[duke]{Department of Mechanical Engineering and Materials Science, Duke University, Durham, NC, 27708, United State of America}

\begin{abstract}
Boron Carbide exhibits a broad composition range, implying a degree of intrinsic substitutional disorder. While the observed phase has rhombohedral symmetry (space group $R\bar{3}m$), the enthalpy minimizing structure has lower, monoclinic, symmetry (space group $Cm$). The crystallographic primitive cell consists of a 12-atom icosahedron placed at the vertex of a rhombohedral lattice, together with a 3-atom chain along the 3-fold axis. In the limit of high carbon content, approaching 20\% carbon, the icosahedra are usually of type B$_{11}$C$^p$, where the $p$ indicates the carbon resides on a polar site, while the chains are of type C-B-C.  We establish an atomic interaction model for this composition limit, fit to density functional theory total energies, that allows us to investigate the substitutional disorder using Monte Carlo simulations augmented by multiple histogram analysis.  We find that the low temperature monoclinic $Cm$ structure disorders through a pair of phase transitions, first via a 3-state Potts-like transition to space group $R3m$, then via an Ising-like transition to the experimentally observed $R\bar{3}m$ symmetry. The $R3m$ and $Cm$ phases are electrically polarized, while the high temperature $R\bar{3}m$ phase is nonpolar.
\end{abstract}


\end{frontmatter}

\section{Introduction}

The phase diagram of boron carbide is not precisely known, with both qualitative and quantitative discrepancies among the different research groups~\cite{Samsonov58,Ekbom81,Beauvy83,Schwetz91,Okamoto92,Domnich11,Rogl14}.  The most widely accepted diagram of Schwetz~\cite{Schwetz91,Okamoto92} displays a single boron carbide phase at temperature above 1000C, coexisting with elemental boron and graphite. The carbon concentration covers the range 9$\%$-19.2$\%$ carbon, falling notably short of the 20\% carbon fraction at which the electron count is believed to be optimal~\cite{Longuet-Higgins55,Lipscomb81,Balakrishnarajan07}.  More interestingly, the nearly temperature independent behavior of the phase boundaries are thermodynamically improbable, and the broad composition range suggests substitutional disorder at 0K, in apparent violation of the 3rd law.  Since so much remains unknown, and experiment can only be assured of reaching equilibrium at high temperature, theoretical calculation offers hope for resolving the behavior at lower temperatures, in addition to interpreting the disorder at high temperatures.

\begin{figure}[ht]
\vskip 0.5cm
\centering\includegraphics[width=0.8\linewidth]{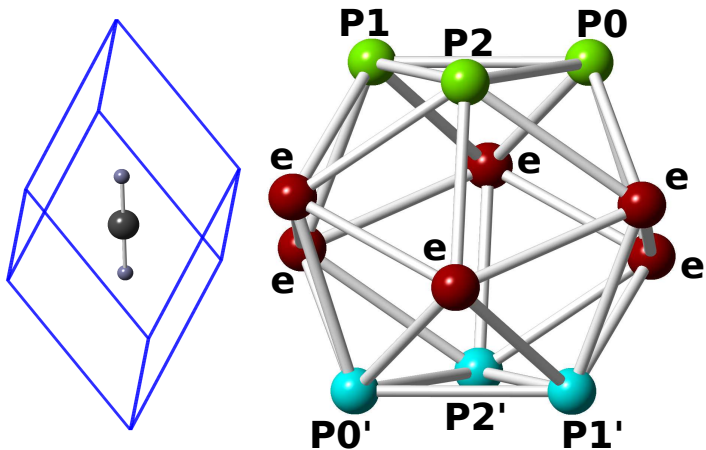}
\caption{Primitive cell of boron carbide showing C-B-C chain at center along the 3-fold axis. The icosahedron (not to scale) occupies the cell vertex. Equatorial sites of the icosahedron are shown in red, labeled ``e'', the north polar sites are shown in green and labeled $p_0$, $p_1$ and $p_2$, while the south polar sites are shown in cyan and labeled as $p_0'$, $p_1'$ and $p_2'$.}
\label{fig:structure}
\end{figure}

As determined crystallographically \cite{GWill76,GWill79,Kwei96}, boron carbide has a 15-atom primitive cell, consisting of an icosahedron and 3-atom chain, in a rhombohedral lattice with symmetry $R\bar{3}m$.  At 20$\%$ carbon a proposed B$_4$C structure featured pure boron icosahedra B$_{12}$ with a C-C-C chain~\cite{Clark43}. Although this structure exhibits $R\bar{3}m$ symmetry, later experimental work \cite{Kwei96,Schmechel00} suggested that the icosahedron should be B$_{11}$C instead of B$_{12}$ and the chain should be C-B-C instead of C-C-C. For other compositions, the icosahedra can be B$_{12}$, B$_{11}$C, or even B$_{10}$C$_2$ (the bi-polar defect~\cite{Mauri01}), and the chain can be C-B-C, C-B-B, B-B$_2$-B~\cite{Yakel,Shirai14} or B-V-B (V means vacancy). Fig.~\ref{fig:structure} illustrates the rhombohedral cell, the C-B-C chain and the icosahedron.  The 12 icosahedral sites are categorized into 2 classes: equatorial and polar.  We further classify the polar sites into north and south.  Icosahedra are connected along edges of the rhombohedral lattice, which pass through the polar sites.

Density functional theory studies of a large number of possible arrangements of boron and carbon atoms~\cite{Mauri01,Widom12,Bylander90} identified four stable phases: pure $\beta$-Boron, rhombohedral B$_{13}$C$_2$, monoclinic B$_4$C, and graphite. The stable rhombohedral phase consists of B$_{12}$ icosahedra with C-B-C chains, giving full rhombohedral symmetry $R\bar{3}m$, while the stable monoclinic phase has B$_{11}C^p$ icosahedra and C-B-C chains.  Carbon occupies the same polar site (e.g. $p_0$) in every icosahedron, resulting in symmetry $Cm$.  Introducing disorder in the occupation of polar sites (i.e. randomly choosing one polar site to occupy with carbon) can restore $R\bar{3}m$ symmetry.  In general, we define site occupations $m_i$ ($i=0, 1, 2, 0', 1', 2'$) corresponding to the mean occupation of the sites $p_i$.  The experimentally observed phase has all $m_i=1/6$.  We call this orientational disorder.  The the stable monoclinic phase has one large order parameter (e.g. $m_0\sim 1$) and the remaining $m_i\sim 0$, which we call orientational order.  Hence it was proposed~\cite{Huhn12} that a temperature-driven order-disorder phase transition is responsible for the high symmetry seen in experiments that are likely in equilibrium only at high temperature.

According to Landau's theory of phase transitions~\cite{Wooten08}, the space groups of structures linked by continuous or at most weakly first order phase transitions should obey group-subgroup relationships. Additionally, the subgroup should be maximal, again provided the transition is continuous or at most weakly first order.  Typically the high temperature phase possesses the higher symmetry, as this permits a higher entropy.  In regard to boron carbide, the high temperature phase has space group $R\bar{3}m$ (group \#166) and the low temperature phase has group $Cm$ (group \#8).  However, $Cm$ is not a maximal subgroup of $R\bar{3}m$, suggesting the possible existence of an intermediate phase.  Two sequences of transitions obey the maximality requirement:
$R\bar{3}m\rightarrow R3m\rightarrow Cm$ and $R\bar{3}m\rightarrow C2/m\rightarrow Cm$~\cite{Hahn84}.  The two corresponding intermediate symmetries $R3m$ and $C2/m$ have space group numbers \#160 and \#12, respectively.

In terms of the distribution of carbon atoms on icosahedra, $R3m$ breaks the inversion symmetry, and hence corresponds to occupying one pole (e.g. the north pole) more heavily than the other, so that $m_i\sim 1/3$ ($i=0, 1, 2$) with the remaining $m_i\sim 0$.  In contrast, $C2/m$ breaks the 3-fold rotational symmetry but preserves inversion.  Thus the carbon atoms preferentially occupy a pair of diametrically opposite polar sites, e.g. $m_0=m_0'\sim 1/2$ while the remaining $m_i\sim 0$.

Since the carbon atom draws charge from surrounding borons, the phases that break inversion symmetry possess an electric dipole moment.  Hence we name the $R3m$ state ``polar''.  The possibility of a polar phase was independently suggested recently~\cite{Ektarawong14}.  We name the $Cm$ state ``tilted polar'' because the broken 3-fold symmetry creates a component of polarization in the $xy$ plane.  Although it lacks a net dipole moment, we name the $C2/m$ state ``bipolar'' because it is reminiscent of the bipolar defect~\cite{Mauri01}. Finally, we name the the high symmetry phase $R\bar{3}m$ ``nonpolar''.

To identify phase transitions, and to determine which symmetry-breaking sequence occurs, we perform Monte Carlo simulations.  Strictly speaking, phase transitions occur only in the thermodynamic limit of large system size, which is beyond the reach of density functional theory calculations. Hence we construct a classical interatomic interaction model, with parameters fit to density functional theory energies.  For simplicity we consider only the high carbon limit where every icosahedron contains a single polar carbon (i.e. we essentially project the composition range onto the $x_C=0.2$ line).  We analyze our simulation results with the aid of the multiple histogram technique~\cite{Swendsen88,Swendsen89}.  In the end we indeed discover a sequence of two phase transitions.  One arises from the breaking of 3-fold symmetry linking $Cm$ to $R3m$ that is first order, similar to the 3-state Potts model~\cite{Potts52} in three dimensions.  The other corresponds to the breaking of inversion symmetry linking $R3m$ to $R\bar{3}m$ that is in the Ising universality class.

\section{Methods}

\subsection{Pair interaction model}

Given that every primitive cell contains a B$_{11}$C$^p$ icosahedron and a C-B-C chain, the configuration can be uniquely specified by assigning a 6-state variable $\sigma$ to each cell, corresponding to which of the six polar sites holds the carbon atom.  The relaxed total energy of a specific configuration can be expressed through a cluster expansion in terms of pairwise, triplet and higher-order interactions~\cite{Sanchez84,Walle02} of these variables.  As shown below, truncating at the level of pair interactions provides sufficient accuracy for present purposes.  Further, we observe that symmetry-inequivalent pairs are in nearly one-to-one correspondence with the inter-carbon separation $R_{ij}=|\bR_i-\bR_j|$ where the $\bR_i$ are the initial positions prior to relaxation and belong to a discrete set of fixed possible values $\{R_k\}$, arranged in order of increasing length.  Note that we need not concern ourselves with interactions of polar carbons with chain carbons, as the number of such pairwise interactions is conserved across configurations.  Thus our total energy can be expressed as
\begin{equation}
\label{eq:bondmodel}
E(N_1,\dots,N_m)=E_0+\sum_{i=1}^m a_k N_k
\end{equation}
where $N_k$ is the number of intercarbon separations of length $R_k$, and $m$ is the number of such separations we choose to treat in our model.

We use the density functional theory-based Vienna ab initio simulation package (VASP)~\cite{Kresse93,Kresse94,Kresse961,Kresse962} to calculate the total energies within the projector augmented wave (PAW)~\cite{Blochl94,Kresse99} method utilizing the PBE generalized gradient approximation~\cite{Perdew96,Perdew97} as the exchange-correlation functional.  We construct a variety of 2x2x2 and 3x3x3 supercells, which we relax with increasing $k$-point meshes until convergence is reached at the level of 0.1 meV/atom, holding the plane-wave energy cutoff fixed at 400 eV.

Using our DFT energies, we fit the $m+1$ parameters $E_0$ and $\{a_i\}$ in our bond interaction model Eq.~(\ref{eq:bondmodel}), increasing $m$ until we are satisfied with the quality of the fit, at $m=10$.  The shortest bond included has length $R_1=1.732$~\AA, corresponding to carbons at polar sites joined by an intericosahedral bond.  For example, $p_0$ and $p_0'$ carbons on icosahedra joined by a bond in the $p_0$ direction.  This bond has the largest strength, with $a_1=1.126$ eV.  Bond strength rapidly diminishes with separation $R$.  Our longest bonds included are $R_9=5.174$~\AA~ (the lattice constant separating neighboring rhombohedral vertices) and $R_{10}=5.465$~\AA~ (the second neighbor rhombohedral vertex separation).

\begin{figure}[htbp]
\vskip 1.0cm
\centering
\includegraphics[width=0.8\linewidth]{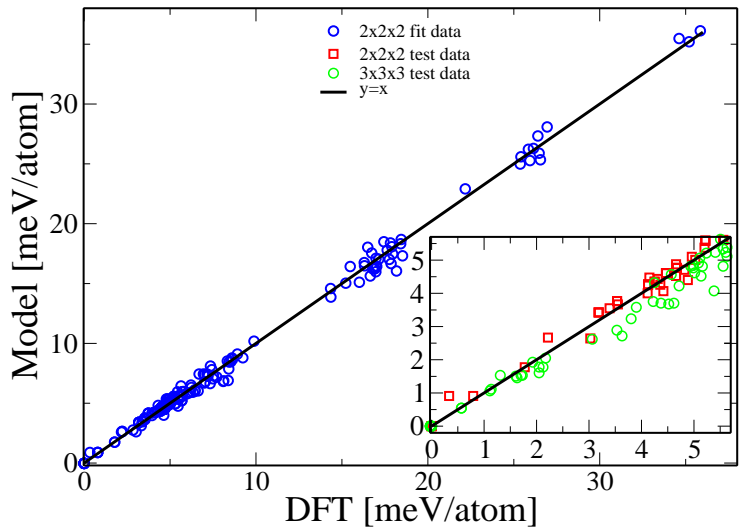}
\caption{Fit of bond interaction model to DFT-calculated total energies in $2\times2\times2$ supercells. Inset shows cross validation check of $2\times2\times2$ supercells transferability to $3\times3\times3$ supercells.}
\label{fig:fitting}
\end{figure}

Our fitting procedure minimizes the mean-square deviation of model energy from calculated DFT energy, supplemented by a small contribution from the $L_1$ norm of the set of coefficients $\{a_k\}$.  Including the $L_1$ norm regularizes the expansion in a manner similar to compressive sensing~\cite{Hart13,Wakin08} and improves transferability to larger cell sizes.  For our fitted data set of 188 independent 2x2x2 supercells (see Fig.~\ref{fig:fitting}) we obtain an RMS error of 0.474meV/atom.  Checking our fit on a different set of 47 $2\times2\times2$ supercells with energy below 5.7 meV/atom we obtain RMS error of 0.233meV/atom, while checking on 57 $3\times3\times3$ supercells yielded RMS error of 0.405meV/atom.  Note that 5.7 meV/atom corresponds to $15\times 5.7=86$ meV/cell corresponding to $kT$ per degree of freedom at T=1000K.

\subsection{Symmetry and order parameters}

Within Landau theory, each possible symmetry breaking is quantified by an order parameter whose transformation properties match irreducible representations of the parent symmetry group.  Space group $R\bar{3}m$ contains point group $D_{3d}$, which is the symmetry group of the triangular antiprism formed by the six polar sites of the icosahedron.  Important elements include 3-fold rotation about the $z$-axis, reflection in a vertical plane containing the $z$-axis, and inversion through the center.

The longitudinal polarization
\begin{equation}
P_z=m_0+m_1+m_2-m_0'-m_1'-m_2'
\end{equation}
transforms as the one dimensional irreducible representation $A_{2u}$, which breaks inversion symmetry, while preserving rotation and reflection, and hence is suitable for characterizing the transition $R\bar{3}m\rightarrow R3m$.  The pair of functions
\begin{equation}
P_{xz}=(m_0-m_0')+\frac{1}{2}(m_1'+m_2'-m_1-m_2), ~~~
P_{yz}={\sqrt{3}\over 2}(m_1+m_2'-m_1'-m_2)
\end{equation}
transform as the two dimensional irreducible representation $E_g$, which additionally breaks both rotational symmetry, and hence characterizes the further transition $R3m\rightarrow Cm$.  Since we will not care which specific orientation is selected at low temperature, we take the norm of the two dimensional representation, and define $P_{xy}=\sqrt{P_{xz}^2+P_{yz}^2}$.  Although we shall not need it, we note that the functions
\begin{equation}
P_x=(m_0+m_0')-\frac{1}{2}(m_1+m_1'+m_2+m_2'), ~~~
P_y={\sqrt{3}\over 2}(m_1+m_1'-m_2-m_2'),
\end{equation}
which transform as the irrep $E_u$, characterize the transformation $R\bar{3}m\rightarrow C2/m$.

As examples of the use of these order parameters, consider a fully disordered nonpolar state of symmetry $R\bar{3}m$ in which all $m_i=1/6$.  All the above order parameters vanish in this state.  Now let $m_i=1/3$ while $m_i'=0$, and note that $P_z=1$, while $P_{xz}=P_{yz}=P_x=P_y=0$, so that $P_z$ indeed characterizes the polar state $R3m$.  Completing the symmetry breaking so that $m_0=1$ while all others vanish, we have both $P_z=1$ and $P_{xz}=1$, so the state is both polar and tilted, with symmetry $Cm$.  Finally, take $m_0=m_0'=1/2$ and all other $m_i$ and $m_i'=0$ and note that $P_x\ne 0$, while $P_z=P_{xz}=P_{yz}=0$, as expected for the bipolar state of symmetry $C2/m$.

\subsection{Monte Carlo simulation and multi-histogram method}

We perform conventional Metropolis Monte Carlo simulations in $L\times L\times L$ supercells of the rhombohedral primitive cell, with $L$ ranging from 3 to 12.  Our basic move is a ``rotation'' in which we randomly select an icosahedron, then randomly displace the carbon from its current polar site to a randomly chosen alternate polar site.  The move is then accepted or rejected according to the Boltzmann factor for the energy change $\Delta E$.  Following an equilibration period, we begin recording the total energy $E$ and the occupations $m_i$ ($i=0, 1, 2, 0', 1', 2'$) of the polar sites for each subsequent configuration.  After a run at one temperature is completed, we take the final configuration as the initial configuration for another run at a nearby temperature.

\begin{figure}[ht]
\vskip 0.5cm
\centering\includegraphics[trim=0 0 0 3,clip,width=0.8\linewidth]{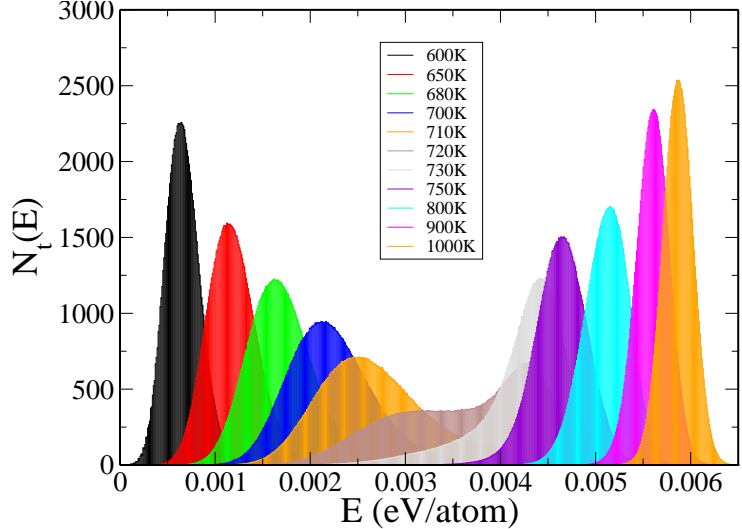}
\caption{Multiple histograms for the 8x8x8 supercell.  The ground state $Cm$ configuration is taken as the zero of energy.}
\label{fig:histograms8}
\end{figure}

At a given simulation temperature $T_s$, a histogram of configuration energies $H_{T_s}(E)$ (see Fig.~\ref{fig:histograms8}) can be converted into a density of states $W(E)=H_{T_s}(E) \exp{(E/kT_s)}$, which is accurate over the energy range that has been well sampled at temperature $T$.  This density of states can be used to calculate the partition function
\begin{equation}
Z(T)=\sum_E W(E) e^{-E/kT}
\end{equation}
which is accurate over a range of temperatures close to $T_s$~\cite{Swendsen88}. The logarithm of $Z(T)$ yields the free energy, and derivatives of the free energy yield other quantities such as internal energy, entropy and specific heat (see Fig.~\ref{fig:cv}).  Alternatively, we may take moments of the energy distribution,
\begin{equation}
\label{eq:avE}
\avg{E^q} = \sum_E W(E) E^q e^{-E/k_BT}.
\end{equation}
The first moment ($q=1$) yields the thermodynamic internal energy, while the fluctuations
\begin{equation}
\label{eq:Cv}
c_v(T)=\frac{\avg{E^2}-\avg{E}^2}{k_BT^2}
\end{equation}
give the specific heat.

Moreover, by combining histograms taken at temperatures chosen so that the tails of the histograms overlap, the density of states can be self-consistently reconstructed~\cite{Swendsen89} so that the free energy becomes accurate over all intervening temperatures.  Inspecting the histograms shown in Fig.~\ref{fig:histograms8}, rapid evolution is apparent with between temperatures 710 and 730K, which can be an indication of a phase transition.  As supercell size increases the histograms narrow, requiring additional simulation temperatures to maintain the degree of overlap seen here.

This notion can be extended to multidimensional histograms in which the density of states is further broken down according to values of order parameters of interest.  For instance, average powers of the longitudinal polarization can be evaluated as
\begin{equation}
\label{eq:avPz}
\avg{|P_z|^q}(T) = \sum_{E, P_z} W(E, P_z) |P_z|^q e^{-E/kT},
\end{equation}
where $W(E, P_z)$ is the joint distribution of energy and longitudinal polarization, and we take the absolute value of $P_z$ because in a well equilibrated simulation both positive and negative values of $P_z$ occur with equal frequency.  The first power gives the mean polarization, while from the first and second powers together we obtain the longitudinal susceptibility
\begin{equation}
\label{eq:chiz}
\chi_z(T)=N\frac{\avg{|P_z|^2} - \avg{|P_z|}^2}{ k_BT},
\end{equation}
where $N$ is the number of atoms. The susceptibility $\chi_{xy}(T)$ is obtained in similar fashion. The units of $\chi_z$ and $\chi_{xy}$ are $eV^{-1}/atom$.

\section{Results and Discussion}

\subsection{Order parameters}

Plotting the order parameters vs. temperature provides a quick way to determine the sequence of phases and transitions.  As Fig.~\ref{fig:4pics} shows, $\avg{|P_z|}$ passes through two regimes of anomalous behavior.  As the supercell size grows, the average longitudinal polarization $\avg{|P_z|}$ vanishes for $T\gtrsim 790$K but approaches to finite values for $T\lesssim 790$K.  At $T=790$K the slope of $\avg{|P_z|}(T)$ diverges.  An even stronger divergence of slope occurs at $T\approx 717$K.  Meanwhile, $\avg{P_{xy}}$ decreases with increasing supercell size for $T\gtrsim 717$K  but approaches finite values for $T\lesssim 717$K. The diverging slope at $T\approx 717$K is consistent with an emerging discontinuity in $\avg{P_{xy}}(T)$.

On the basis of the order parameters, we judge there are three phases separated by two phase transitions.  The high temperature phase has symmetry $R\bar{3}m$ both $P_z$ and $P_{xy}$ vanish. Below 790K a longitudinal polarization grows continuously, and we enter a phase of symmetry $R3m$, having lost inversion symmetry.  Around 717K the polarization suddenly tilts off the $z$-axis and we enter the tilted polar phase of symmetry $Cm$.

\subsection{Specific heat and susceptibility}

Having explored the order parameters, which can be considered as first derivatives of the free energy with respect to applied fields, we now consider the specific heat and susceptibilities.  The specific heat corresponds to a second derivative of free energy with respect to temperature, while the susceptibilities are second derivatives with respect to conjugate fields.  All are evaluated from Monte Carlo data via the fluctuations formulas such as Eqs.~(\ref{eq:Cv}) and~(\ref{eq:chiz}).

\begin{figure}[ht]
\vskip 0.5cm
\centering\includegraphics[trim=0 0 0 3,clip,width=0.8\textwidth]{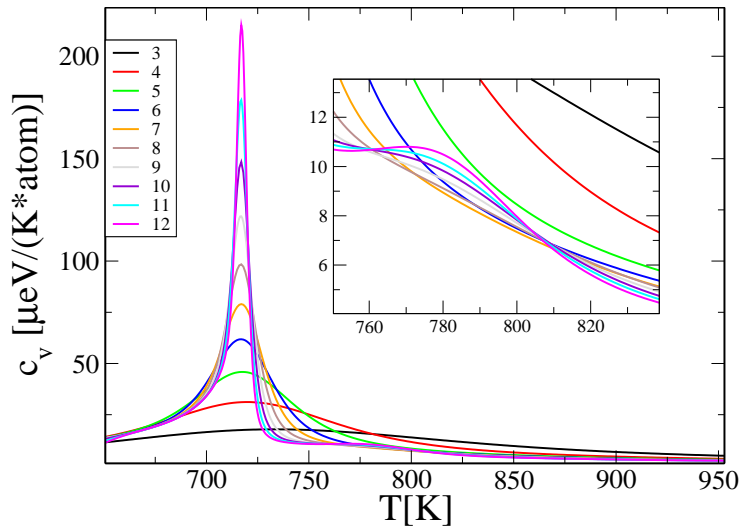}
\caption{Specific heat for 3x3x3 to 12x12x12 supercells.}
\label{fig:cv}
\end{figure}

\begin{sidewaysfigure}[ht]
\centering
\includegraphics[width=1.1\linewidth]{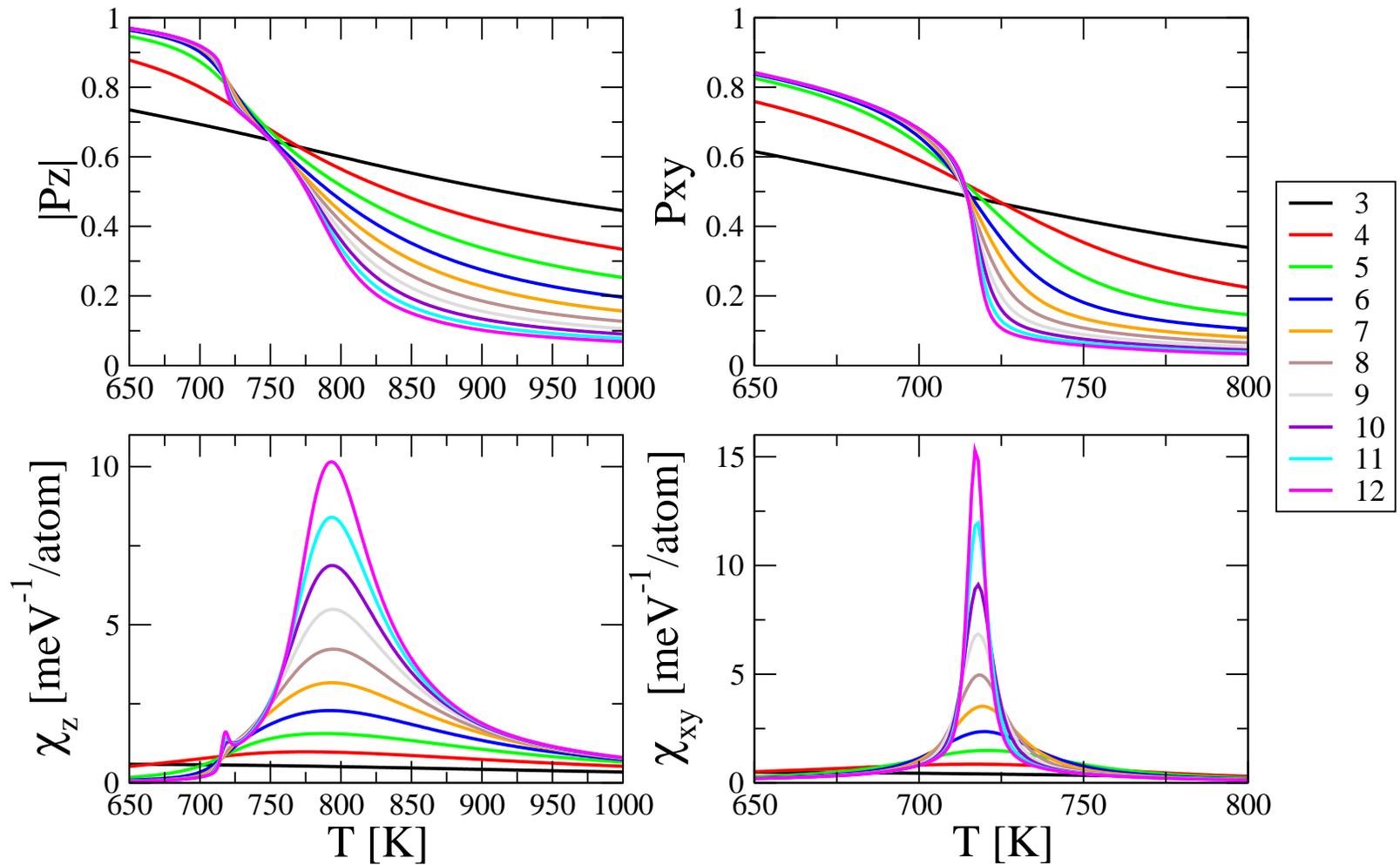}
\caption{Order parameters $|P_z|$ (upper left), $Pxy$ (which is $P_\perp$, upper right) and corresponding susceptibility $\chi_z$ (lower left), $\chi_{xy}$ (lower right).}
\label{fig:4pics}
\end{sidewaysfigure}

Specific heat for a series of increasing supercell sizes is shown in Fig.~\ref{fig:cv}.  In addition to a strong peak around T=717K, a weak peak around T=790K can be seen growing for larger supercell sizes in the inset.  The growing peaks converge to temperatures that roughly correspond to the order parameter anomalies seen above.  Fig.~\ref{fig:4pics} shows the longitudinal and perpendicular (i.e. in-plane) susceptibilities, $\chi_z$ and $\chi_{xy}$ respectively.  Evidentally the high temperature specific heat peak coincides with the peak in $\chi_z$, and hence relates to the fluctuations associated with onset of longitudinal polarization.  Similarly, the low temperature specific heat peak coincides with the peak in $\chi_{xy}$, and hence relates to fluctuations associated with the tilt of polarization off the 3-fold axis.

\subsubsection{Ising-like transition}

As the high temperature transition from $R\bar{3}m$ to $R3m$ coincides with a breaking of inversion symmetry, we expect the transition to be in the universality class of the three-dimensional Ising model.  Some associated critical exponents are $\alpha=0.110$ (specific heat), $\gamma=1.2372$ (susceptibility) and $\nu=0.6301$ (correlation length)~\cite{Pelissetto02}.  Applying finite size scaling theory~\cite{Landau05}, we note that the specific heat peak height should diverge with increasing supercell size as $L^{\alpha/\nu}$, where $\alpha/\nu=0.175$.  The small value of this exponent explains the weak divergence seen around 790K in Fig.~\ref{fig:cv}.  Similarly the susceptibility $\chi_z$ should diverge as $L^{\gamma/\nu}$, with $\gamma/\nu=1.963$.

\begin{figure}[ht]
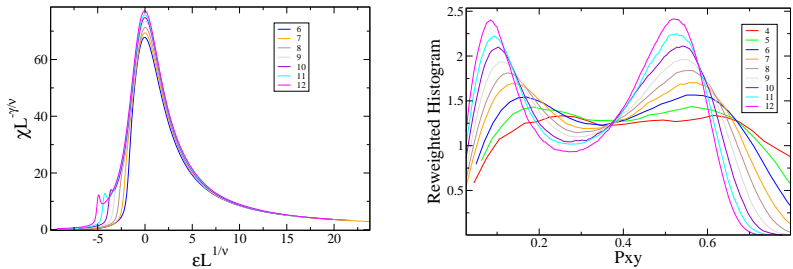

\vskip 0.5cm
\centering
\includegraphics[width=0.4\linewidth]{Isingscaling.eps}
\hspace{0.2in}
\includegraphics[width=0.4\textwidth]{rwthistpxy.eps}
\caption{Validation of universality classes. (left) Ising scaling function for $\chi_z$; (right) Lee-Kosterlitz histograms of $P_\perp$.}
\label{fig:Isingscale}
\end{figure}

Validation of the size- and temperature-dependence of $\chi_z$ requires a finite-size scaling collapse of axes, plotting the scaled susceptibility $\chi_z/L^{\gamma/\nu}$ as a function of an expanded temperature scale $\epsilon L^{1/\nu}$ where $\epsilon=(T-T_c)/T_c$.  When plotted in this manner as seen in Fig.~\ref{fig:Isingscale} the finite size susceptibilities converge to a common scaling function $\chi_0$, supporting the proposed Ising universality class of this continuous phase transition as well as yielding an improved estimate for $T_c=793.7$K.

\subsubsection{3-state Potts-like transition}

Once a direction for longitudinal polarization has been chosen at the high temperature Ising-like transition (e.g. north), the remaining orientational ordering requires selecting a particular in-plane direction (e.g. $i=0, 1$ or 2), resulting in a breaking of 3-fold rotational symmetry.  Thus we expect the low temperature transition to be in the universality class of the 3-state Potts model.  In three dimensions this transition is expected to be weakly first order~\cite{Wu82}.

Because the order parameter jumps discontinuously at a first order transition, the fluctuations per atom of energy and polarization should grow proportionally to the number of atoms, i.e. as $L^3$.  When peak heights of $c_v$ and $\chi_{xy}$ are plotted on a log-log plot vs. $L$, we expect a straight line in the asymptotic limit of large $L$ whose slope should be 3.  Unfortunately, our largest supercell size $L=12$ has not yet reached this limit, with the slopes around 2 and 2.8 seen for $c_v$ and $\chi_{xy}$ respectively, clearly tending to increase with $L$.

The Lee-Kosterlitz criterion~\cite{Landau05,Lee90} is an alternative method to confirm a first order transition.  Because the two coexisting phases exhibit finite differences in properties such as energy and polarization, probability distributions of such properties should then be bimodal, with each peak sharpening as system size grows.  Fig.~\ref{fig:Isingscale} illustrates this distribution for $P_{xy}$.  This distribution is obtained by marginalizing the joint energy and polarization histogram $H_{T_s}(E,P_{xy})$ over energy, then reweighting with the factor $\exp(E/k_BT_e-E/k_BT_s)$, where the temperature $T_e$ is chosen so as to make the heights of the two peaks equal.  Clearly the distributions of polarization illustrate coexistence of a state with $P_{xy}=0$ and a state with $P_{xy}\sim 0.5$.  Thus we conclude the transition is first order, as expected for symmetry-breaking of the 3-state Potts type in three dimensions.

\subsection{Electric dipole moments}

Because of the charge imbalance created by the polar carbons, the polar and tilted polar states must exhibit electric dipole moments, while the nonpolar state does not.  We constructed specific representative structures of each of the three phases to calculate these dipole moments. Taking a single hexagonal unit cell containing three primitive cells, we constructed the tilted polar $Cm$ state by placing carbon at each of the three $p_0'$ sites.  We then constructed a polar $R3m$ state by placing the polar carbon at $p_0'$ in one cell, $p_1'$ in another and $p_2'$ in the third. Finally we took a $2\times1\times1$ supercell of the hexagonal unit cell, and inverted the polarization of every second carbon (e.g. we replace the $p_0'$ in the first hexagonal cell by $p_0$, and did the same for $p_1'$ and $p_2'$ in the second and third hexagonal cell), resulting in a nonpolar state that locally resembles $R\bar{3}m$.  Electric dipole moments as calculated by VASP are given in Table~\ref{tab:dipole}.

\begin{table}
\centering
\begin{tabular}{|c|c|l|l|l|}
\hline
Phase & symmetry group & $p_x$ & $p_y$ & $p_z$ \\
\hline
tilted polar & $Cm$ & -0.63437 & 0.36626 & 1.13 \\
\hline
polar & $R3m$ & 0.00 & 0.00 & 1.11 \\
\hline
nonpolar & $R\bar{3}m$ & 0.00 & 0.00 & 0.00 \\
\hline
\end{tabular}
\caption{Electric dipole moments of the 3 phases (units are $e$\AA, where $e$ is the magnitude of the charge on an electron. The polar phase has p0',p1',p2' carbons, resulting in dipole moment along +z direction. For $Cm$ phase, the projection onto xy-plane of dipole moment is along the projection of the vector $\bP0'$ (1.894,-1.093,-2.666) from the center of icosahedra to p0'.}
\label{tab:dipole}
\end{table}

\section{Conclusion}

We construct an artificial model inspired by boron carbide by placing an orientational degree of freedom at the vertices of a rhombohedral lattice, mimicking the distribution of carbon sites among polar vertices of B$_{11}$C$^p$ icosahedra.  Because this model is restricted to 20\% carbon it cannot capture the broad composition range of true boron carbide, but it can reveal orientational order and disorder similar to what might be seen in experiment.  A pairwise interaction model counting bonds of specific type between polar carbons fits well to density functional theory total energies, and is transferable between supercells of differing sizes.

Monte Carlo simulations utilizing this model reveal three distinct phases separated by a pair of phase transitions.  The high temperature phase has symmetry group $R\bar{3}m$, similar to what is observed experimentally in boron carbide.  As temperature falls to 790K, inversion symmetry is lost via a continuous phase transition, resulting in a polarized state of symmetry $R3m$, which is a maximal subgroup of $R\bar{3}m$.  The possible existence of such a state was independently suggested~\cite{Ektarawong14}, although at a much higher temperature.  Finally, as temperature falls below 717K, 3-fold rotational symmetry is broken via a first order transition, resulting in a tilted polar phase of monoclinic symmetry $Cm$, which is a maximal subgroup of $R3m$.  When fully orientationally ordered, this state matches the previously known ground state of B$_4$C~\cite{Mauri01,Widom12,Bylander90}.

The universality classes of each transition follow the expectations based on the type of symmetry breaking.  The continuous 790K transition, which breaks inversion symmetry, is shown to fall in the Ising universality class because of the finite size scaling collapse of longitudinal susceptibility $\chi_z$ as shown in Fig.~\ref{fig:Isingscale}.  The 717K transition, which breaks 3-fold rotation symmetry, is shown by the Lee-Kosterlitz criterion to be weakly first order, consistent with expectations for the 3-state Potts universality class.

\section{Acknowledgements}
We thank Robert H. Swendsen, David P. Landau and James P. Sethna for helpful discussions.

\section{References}
\bibliographystyle{./elsarticle-num}

\begin{thebibliography}{10}
\expandafter\ifx\csname url\endcsname\relax
  \def\url#1{\texttt{#1}}\fi
\expandafter\ifx\csname urlprefix\endcsname\relax\def\urlprefix{URL }\fi
\expandafter\ifx\csname href\endcsname\relax
  \def\href#1#2{#2} \def\path#1{#1}\fi

\bibitem{Samsonov58}
Samsonov G.V., The present state of the investigation of the boron-carbon diagram, Zh. Fiz. Khim., Vol. 32, 1958, p 2424-2429 (in Russian).

\bibitem{Ekbom81}
Ekbom, Lars B., Amundin, Carl Olof, Microstructural evaluation of sintered boron carbides with different compositions, Science of Ceramics, 11, p 237-243.

\bibitem{Beauvy83}
Beauvy M., Stoichiometric limits of carbon-rich boron carbide phases, J. Less-Common Met., Vol. 90, 1983, p 169-175.
  
\bibitem{Schwetz91}
K.~A. Schwetz, P.~Karduck, Investigations of the boron-carbon system with the aid of electron probe microanalysis, J. Less Common Met. 175 (1991) 1--100.

\bibitem{Okamoto92}
H.~Okamoto, \uppercase{B}-\uppercase{C} (boron-carbon), J. Phase Equil. 13 (1992) 436.

\bibitem{Domnich11}
Domnich, V., Reynaud, S., Haber, R.A., Chhowalla, M. Boron carbide: Structure, properties, and stability under stress, Journal of the American Ceramic Society (2011), 94 (11), p 3605-3628.  

\bibitem{Rogl14}
Peter F. Rogl, Jan V\v re\v st'\' al, Takaho Tanaka and Satoshi Takenouchi, The B-rich side of the B-C phase diagram, Calphad 44 (2014) 3-9.

\bibitem{Longuet-Higgins55}
H. C. Longuet-Higgins and M. de V. Roberts, The electronic structure of an icosahedron of boron atoms, Proc. R. Soc. Lond. A 1955 230, 110-119.

\bibitem{Lipscomb81}
W. N. Lipscomb, Borides and boranes, J. Less-Common Metals 82 (1981) 1-20.

\bibitem{Balakrishnarajan07}
Musiri M. Balakrishnarajan,z Pattath D. Pancharatna and Roald Hoffmann,
Structure and bonding in boron carbide: The invincibility of imperfections,
New J. Chem. 31 (2007) 473-85.

\bibitem{GWill76}
G. Will, K. H. Kossobutzki, An x-ray structure analysis of boron carbide, b13c2, J. Less-Common Met. 44 (1976) 87.

\bibitem{GWill79}
G. Will, A. Kirfel, A. Gupta, E. Amberger, Electron density and bonding in b13c2, J. Less-Common Met. 67 (1979) 19-29.

\bibitem{Kwei96}
G. H. Kwei, B. Morosin, Structures of the boron-rich boron carbides from neutron powder diffraction: Implications for the nature of the intericosahedral chains, J. Phys. Chem. 100 (1996) 8031-9.

\bibitem{Clark43}
H. K. Clark, J. L. Hoard, The crystal structure of boron carbide, J. Am. Chem. Soc. 65 (1943) 2115-9.

\bibitem{Schmechel00}
R. Schmechel, H. Werheit, Structural defects of some icosahedral boron-rich solids and their correlation with the electronic properties, J. Solid State Chem. 154 (2000) 61-7.

\bibitem{Mauri01}
F. Mauri, N. Vast, C. J. Pickard, Atomic structure of icosahedral b4c boron
carbide from a first principles analysis of nmr spectra, Phys. Rev. Lett. 87
(2001) 085506.

\bibitem{Yakel}
Yakel, H. L., The crystal structure of a boron-rich boron carbide, Acta Crystallogr. B 31, 1797 (1975).

\bibitem{Shirai14}
Koun Shirai, Kyohei Sakuma and Naoki Uemura, Theoretical study of the structure of boron carbide B$_13$C$_2$, Phys. Rev. B 90 (2014).

\bibitem{Widom12}
M. Widom and W. Huhn, Prediction of Orientational Phase Transition in Boron Carbide, Solid State Sciences 14, 1648 (2012), ISSN 1293-2558.

\bibitem{Bylander90}
D. M. Bylander, L. Kleinman, S. Lee, Self-consistent calculations of the
energy bands and bonding properties of b12c3, Phys. Rev. B 42 (1990)
13941403.

\bibitem{Huhn12}
W.P. Huhn and M. Widom, A free energy model of boron carbide, J. Stat. Phys. 150 (2012) 432-41.

\bibitem{Wooten08}
El-Batanouny M, Wooten F. Symmetry and condensed matter physics: a computational approach [M]. Cambridge University Press, 2008.

\bibitem{Hahn84}
Hahn, T, and Paufler, P. International tables for crystallography, Vol. A. space-group symmetry D. REIDEL Publ. 1984.

\bibitem{Ektarawong14}
A. Ektarawong, S. I. Simak, L. Hultman, J. Birch, and B. Alling, First-principles study of configurational disorder in B4C using a superatom-special quasirandom structure method, Phys. Rev. B 90 (2014) 024204.

\bibitem{Swendsen88}
A. M. Ferrenberg and R. H. Swendsen, Phys. Rev. Lett. 61, 2635 (1988).

\bibitem{Swendsen89}
A. M. Ferrenberg and R. H. Swendsen, Phys. Rev. Lett. 63, 1195 (1989).

\bibitem{Potts52}
R. B. Potts, Some generalized order-disorder transformations, Mathematical Proceedings of the Cambridge Philosophical Society. Cambridge University Press, 1952, 48(01): 106-109.

\bibitem{Sanchez84}
J. M. Sanchez, F. Ducastelle and D. Gratias, Generalized cluster description of multicomponent systems, Physica 128A (1984) 334-50.

\bibitem{Walle02}
A. van de Walle, M. Asta and G. Ceder, The alloy theoretic automated toolkit: A user guide, Calphad 26 (2002) 539-53.

\bibitem{Kresse93}
G. Kresse and J. Hafner. Ab initio molecular dynamics for liquid metals. Phys. Rev. B, 47:558, 1993.

\bibitem{Kresse94}
G. Kresse and J. Hafner. Ab initio molecular-dynamics simulation of the liquid-metal-amorphous-semiconductor transition in germanium. Phys. Rev. B, 49:14251, 1994.

\bibitem{Kresse961}
G. Kresse and J. Furtmuller. Efficiency of ab-initio total energy calculations for metals and semiconductors using a plane-wave basis set. Comput. Mat. Sci., 6:15, 1996.

\bibitem{Kresse962}
G. Kresse and J. Furthmuller. Efficient iterative schemes for ab initio total-energy calculations using a plane-wave basis set. Phys. Rev. B, 54:11169, 1996.

\bibitem{Blochl94}
P. E. Blochl. Projector augmented-wave method. Phys. Rev. B, 50:17953, 1994.

\bibitem{Kresse99}
G. Kresse and D. Joubert. From ultrasoft pseudopotentials to the projector augmented-wave method. Phys. Rev. B, 59:1758, 1999.

\bibitem{Perdew96}
J. P. Perdew, K. Burke, and M. Ernzerhof. Generalized gradient approximation made simple. Phys. Rev. Lett., 77:3865, 1996.

\bibitem{Perdew97}
J.P. Perdew, J.A. Chevary, S.H. Vosko, K.A. Jackson, M.R. Pederson, D.J. Singh, and C. Fiolhais. Erratum: Atoms, molecules, solids, and surfaces: Applications of the generalized gradient approximation for exchange and correlation. Phys. Rev. B, 48:4978, 1993.

\bibitem{Hart13}
Nelson L J, Hart G L W, Zhou F, et al. Compressive sensing as a paradigm for building physics models[J]. Physical Review B, 2013, 87(3): 035125.

\bibitem{Wakin08}
Cand\`es E J, Wakin M B. An introduction to compressive sampling[J]. Signal Processing Magazine, IEEE, 2008, 25(2): 21-30.

\bibitem{Pelissetto02}
A. Pelissetto and E. Vicari, Critical phenomena and renormalization-group thoery, Physics Reports 368, 549 (2002).

\bibitem{Landau05}
D. Landau and K. Binder, A Guide to Monte Carlo Simulations in Statistical Physics, Cambridge University Press, New York, NY, USA, 2005, ISBN 0521842387.

\bibitem{Wu82}
F. Y. Wu, The Potts model, Rev. Mod. Phys. 54, 235 (1982).

\bibitem{Lee90}
J. Lee and J. M. Kosterlitz, New numerical method to study phase transitions, Phys. Rev. Lett. 65, 137 (1990).

\end{thebibliography}

\end{document}